JASS *Journal of Astronomy and Space Sciences*

# The Spectral Sharpness Angle of Gamma-ray Bursts

**Hoi-Fung Yu[1,2†], Hendrik J. van Eerten[1], Jochen Greiner[1,2], Re'em Sari[3], P. Narayana Bhat[4], Andreas von Kienlin[1], William S. Paciesas[5], Robert D. Preece[6]**

[1]Max Planck Institute for Extraterrestrial Physics, Garching, Bayern 85748, Germany
[2]Excellence Cluster Universe, Technical University of Munich, Munich, Bayern 80333, Germany
[3]The Hebrew University of Jerusalem, Jerusalem, Israel
[4]Center for Space Plasma and Aeronomic Research, University of Alabama in Huntsville, Huntsville, AL 35748, USA
[5]Universities Space Research Association, Huntsville, AL 35805, USA
[6]Space Science Department, University of Alabama in Huntsville, Huntsville, AL 35809, USA

We explain the results of Yu et al. (2015b) of the novel sharpness angle measurement to a large number of spectra obtained from the *Fermi* gamma-ray burst monitor. The sharpness angle is compared to the values obtained from various representative emission models: blackbody, single-electron synchrotron, synchrotron emission from a Maxwellian or power-law electron distribution. It is found that more than 91% of the high temporally and spectrally resolved spectra are inconsistent with any kind of optically thin synchrotron emission model alone. It is also found that the limiting case, a single temperature Maxwellian synchrotron function, can only contribute up to $58^{+23}_{-18}$% of the peak flux. These results show that even the sharpest but non-realistic case, the single-electron synchrotron function, cannot explain a large fraction of the observed spectra. Since any combination of physically possible synchrotron spectra added together will always further broaden the spectrum, emission mechanisms other than optically thin synchrotron radiation are likely required in a full explanation of the spectral peaks or breaks of the GRB prompt emission phase.

**Keywords**: gamma-rays: stars, gamma-ray burst: general, radiation mechanisms: non-thermal, radiation mechanisms: thermal, methods: data analysis

## 1. INTRODUCTION

Gamma-ray bursts (GRBs) are the most luminous gamma-ray transients ever observed by humankind. There are two different observed phases of electromagnetic emissions, namely the prompt and afterglow phases. The fireball model (Goodman 1986; Rees & Meszaros 1992; Meszaros et al. 1993; Meszaros & Rees 1993; Rees & Meszaros 1994; Tavani 1996; Piran 1999) states that the ejected materials with various Lorentz factors (~ 100 – 1,000) from the central engine form a bipolar jet structure. When the shells of matter of the jet collide, shock waves will be formed and prompt gamma-rays are emitted from the shock-accelerated materials. When the matter shells continue to travel outward, they will eventually interact with the circumburst medium, where similar processes take place and afterglow (from gamma-ray to radio) is emitted.

The emission mechanisms of the prompt and afterglow phases provide information for the physical processes at work. For decades, astrophysicists believe that the prompt emission is generated by synchrotron radiation from electron populations as the afterglow emission. However, observations of the prompt phase are not as conclusive as the afterglow phase since the prompt phase is only observed in gamma-rays for most GRBs, whereas the synchrotron origin of afterglow is validated by multi-wavelength observations of many bursts.

Gamma-ray spectroscopy is the key to investigate the emission mechanism of the prompt phase. Thanks to the gamma-ray burst monitor (GBM, Meegan et al. 2009) onboard the *Fermi* Gamma-ray Space Telescope, high spectral and temporal resolution spectra are obtained for > 2,000 GRBs since its launch in July 2008. The GBM covers a very wide spectral range from 8 keV – 40 MeV by two kinds of detectors, the thallium activated sodium iodide (NaI(Tl)) detectors responsible for lower energy (8 – 900 keV) and the bismuth germanate (BGO) detectors responsible for higher energy (250 keV – 40 MeV).

The study of the sharpness of the synchrotron emission spectrum in comparison to time-resolved spectra of GRBs is a question recently raised by Beloborodov (2013) and Vurm & Beloborodov (2015). In this proceeding we discuss the results presented in Yu et al. (2015b), in which a total of 1,491 time bins with constrained spectral parameters are obtained (see the official time-resolved spectral catalog, Yu et al. 2016). For 1,113 of them, a spectral peak or break is present within the GBM spectral window. Using a novel quantity called the sharpness angle, we are able to quantify the spectral curvature of these 1,113 spectra and comparing to theoretical emission models.

In Section 2, we describe how this novel quantity is constructed. The key results are presented in Section 3. Some of the theoretical implications are discussed in Section 4. The conclusions are given in Section 5. Unless otherwise stated, all errors reported in this paper are given at the $1\sigma$ confidence level.

## 2. SHARPNESS ANGLE DEFINITION

The spectra are re-fit as described in Yu et al. (2015b). We normalize the re-fit model curves in logarithmic space for those 1,113 spectra by setting the peak energy $E_p$ or break energy $E_b$ and the corresponding energy flux $vF_\nu$ at $(x, y)$ = (1, 1). Then, we can construct a triangle $\{(1, 1), (0.1\ E_p, vF_\nu(0.1E_p)), (3.0\ E_p, vF_\nu(3.0E_p))\}$ below the model curve. The sharpness angle $\theta$ is defined as the angle below (1, 1) in this triangle (Fig. 1). By such definition, $\theta$ is an indication of the spectral sharpness and is independent of the actual position of $E_p$, which means they are also independent of redshift.

## 3. RESULTS

Fig. 2 shows the cumulative distribution function (CDF) of the sharpness angles $\theta$ and the distributions of the errors $\sigma_\theta$. The dotted, solid, and dashed black vertical lines indicate the values of $\theta$ for the normalized blackbody, single-electron synchrotron emission function, and synchrotron emission function from a Maxwellian electron distribution, from left to right. It is found that over 35% of the spectra are inconsistent with single-electron synchrotron emission and 91% are inconsistent with synchrotron emission from a Maxwellian electron distribution. The blackbody spectrum is found to be much sharper than any of the observed spectra. Note that the synchrotron emission function from a Maxwellian electron distribution produces one of the sharpest (i.e., narrowest) spectra (see Section 2.2 of Yu et al. 2015b).

Instead of directly propagate the errors on the observed photon counts to the sharpness angles, Monte-Carlo simulations are performed using the $1\sigma$ errors from the best-fit spectral parameters. This is because the spectral peak can only be found and the flux can only be normalized when the counts are convolved with a model and the response matrices, through the official GBM spectral analysis software RMFIT. The errors on the spectral parameters are not necessarily Gaussian distributed, therefore new values are randomly drawn from a uniform probability distribution function which has the largest standard deviation. In this way, our estimation of $\sigma_\theta$ is most conservative. As shown in Fig. 2, the resulting distribution of $\sigma_\theta$ has a median around 5 degrees, too small to affect our conclusions.

It is found that 35% of the 1,113 spectra violate the synchrotron line-of-death (Katz 1994; Tavani 1995; Preece et al. 1998; Crider et al. 1999; Preece et al. 2002), higher than the 20% observed by Gruber et al. (2014) in their peak-flux "P" spectra sample. This implies that a large number of spectra are still consistent with the line-of-death. However, we find that in the 65% of spectra that do not violate the

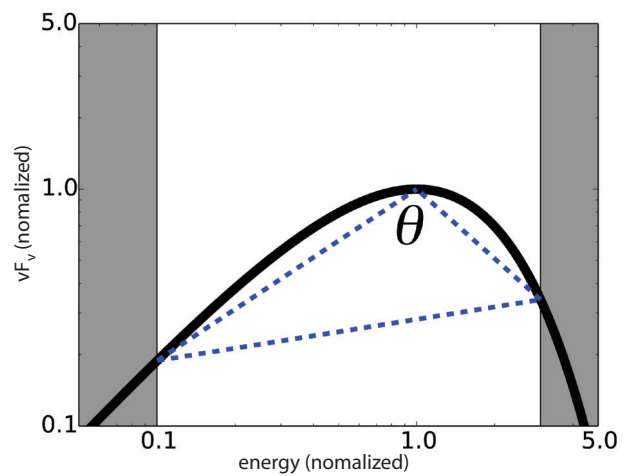

y-axis label: $vF_\nu$ (nomalized)

x-axis label: energy (nomalized)

**Fig. 1.** Illustration of how the triangle is constructed and the sharpness angle $\theta$ is defined. The vertical and horizontal axis are plotted in logarithmic scale in units of normalized $vF_\nu$ flux and photon energy, respectively.

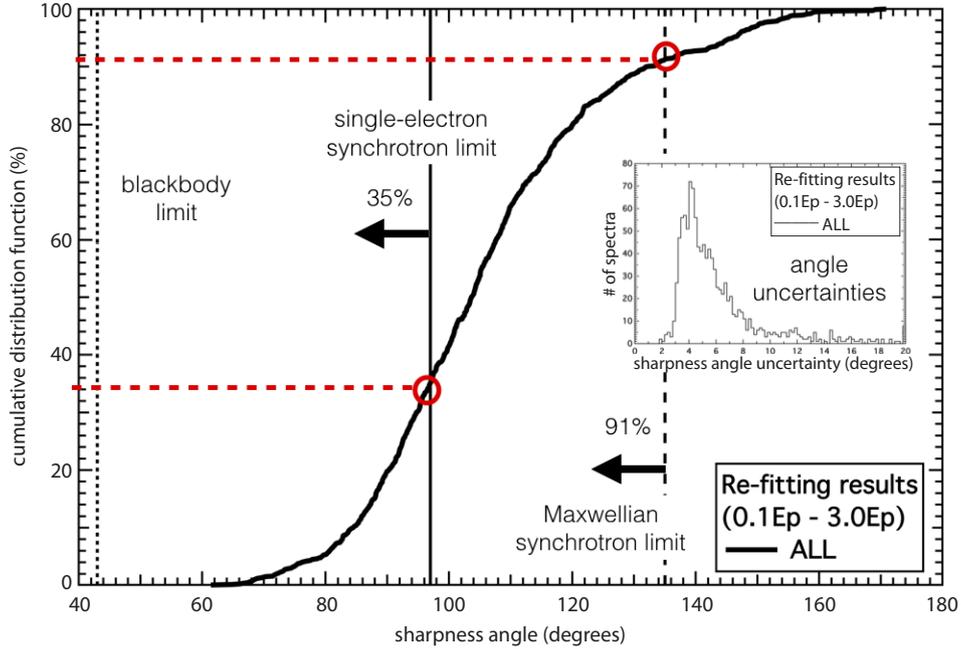

**Fig. 2.** Cumulative distribution functions of $\theta$ and distributions of $\sigma_\theta$. The limits of the normalized blackbody (dotted line), single-electron synchrotron (solid line), and synchrotron with a Maxwellian distribution function (dashed line) are overlaid.

line-of-death, 92% of them violate the Maxwellian limit. This shows that the sharpness angle method can identify many more spectra that are consistent with the line-of-death but are still sharper than what the synchrotron theory predicts. By contrast, of the 35% of spectra that violate the line-of-death, only 10% of them do not violate the Maxwellian limit.

In Fig. 3, we show the distribution of the maximum peak flux allowed by the best-fit model curves to be contributed by the Maxwellian synchrotron function. For the spectra that do not have a peak, we compute this value at the spectral break. A sample spectrum from GRB 101014.175 is plotted in Fig. 4. The normalized Maxwellian synchrotron function was shifted vertically and horizontally until the distance between its value at $x = 1$ and the peak of the fit model is minimized. The advantage of evaluating this value at the peak of the fit model is that it is energy domain independent. It is found that the Maxwellian can only contribute up to $58^{+23}_{-18}$% of the peak flux (solid histogram). Even if the minimum sharpness angles (i.e., the broadest spectra) allowed by the uncertainties in the best-fit parameters are considered, this percentage only slightly increases to $68^{+23}_{-23}$% (dashed histogram).

We select and plot in Fig. 5 the evolution of $\theta$ for 6 example bursts, with the Maxwellian synchrotron limit and the observed light curves overlaid. It can be seen that $\theta$ exhibits various evolutionary trends. These bursts are chosen to show the variety of evolutionary trends in $\theta$:

gradual increase, gradual decrease, fluctuation between the single-electron and Maxwellian limits, small $\theta$ during low emission level and large $\theta$ during high emission level, large $\theta$ during low emission level and small $\theta$ during high emission level, and decrease from above the Maxwellian limit followed by an increase again to above the Maxwellian limit. No general evolutionary trend is found from our burst sample.

## 4. DISCUSSIONS

Our results show that for most GRB prompt emission spectra, an explanation in terms of synchrotron radiation can be problematic. In the internal shocks of GRBs, a single-electron emission function is obviously non-realistic (as there must be multiple electrons in the outflow) and a Maxwellian population drawn from a single temperature is the limiting case. Even this limiting case is already too wide to fit most GRB time-resolved spectra.

For the past two decades, the so-called Band function (Band et al. 1993) has been assumed to be the appropriate mathematical function in fitting most GRB prompt spectra. The fact that most spectra are best fit by the exponential cutoff power-law model (see Yu et al. 2016) shows that the high-energy tail of the prompt spectrum is actually sharper than a Band function would predict. Recently, Ackermann et al. (2012) showed that the Band function's high-energy power laws obtained from GBM spectral fits are too hard for a subsample of GBM bursts with upper limits from the *Fermi* large area telescope (LAT, Atwood et al. 2009). All these results are indicating that the Band function can lead to incorrect interpretation of the data.

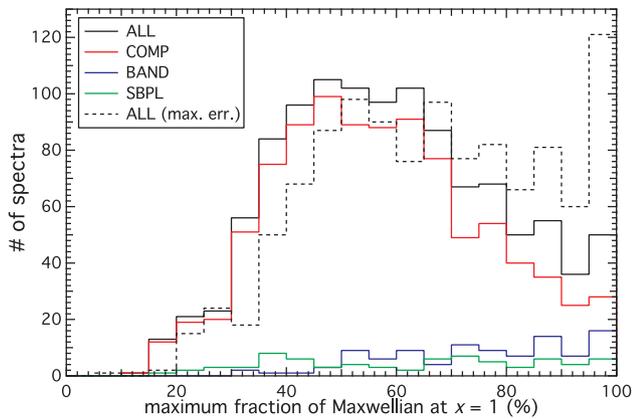

**Fig. 3.** Distribution of the maximum fraction contributed from the Maxwellian synchrotron function at $x = 1$. The solid histograms represent the distributions using the best-fit model parameters, while the dashed histogram shows the minimum allowed sharpness by the uncertainties from the best-fit parameters. Spectra with 100% at $x = 1$ are accumulated in the last bin.

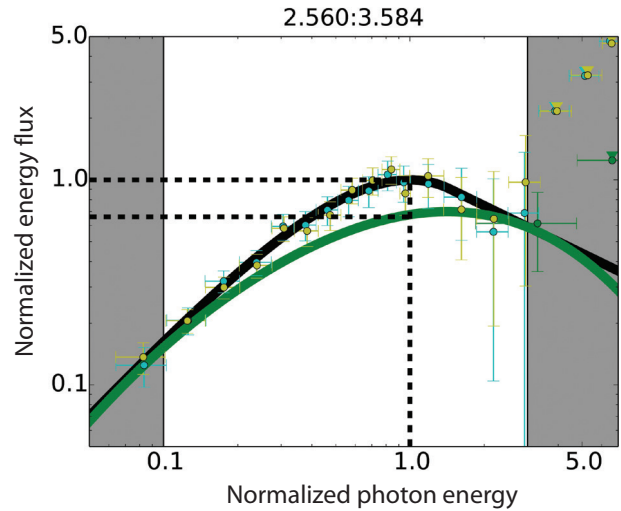

**Fig. 4.** Example spectrum taken from GRB 101014.175 (2.560 − 3.584 sec), showing the maximum contribution to the best-fit model by the Maxwellian synchrotron function, at $x = 1$. The normalized Maxwellian synchrotron (green curve) and the best-fit model (black curve) overlaid. The black dashed lines show the peak position of the best fit model and the relative normalized flux levels. In this particular spectrum, the Maxwellian fraction is about 65% at $x = 1$. Deep green data points are from the BGO detector and the others are from the NaI detectors. Triangles represent upper limits. For display purpose, the bin size has been increased by a factor of 5 − 10 relative to the standard bin size.

Constructing another empirical function to improve upon the Band function fits is very difficult, because the Band function is already very simple mathematically and statistically. Yu et al. (2015a) have shown that a triple power law with sharp breaks constrained according to the synchrotron models could only perform as good as the Band function. In many cases, an extra blackbody is needed to adequately describe the spectral curvature. Recently, for instance, Burgess et al. (2011, 2014) performed physical model fits, and Uhm & Zhang (2014) have done simulations under more realistic physical condition, e.g., a decaying magnetic field. However, without knowledge of the true emission process, it is difficult to formulate a sufficiently well-constrained physical fit function. There may also be multiple emission mechanisms at work, the sum of which forms the observed prompt spectra. The sharpness angle distribution implies that any model based on standard synchrotron theory without additional radiative and/or absorption mechanisms will systematically struggle to capture the prompt spectral curvature.

Recently, Axelsson & Borgonovo (2015) have shown that using the full-width-half-maximum measurement of GRB prompt emission spectra taken from the BATSE 5B GRB spectral catalog (Goldstein et al. 2013) and 4-years *Fermi* GBM GRB time-integrated spectral catalog (Gruber et al. 2014), a significant fraction of bursts (78% for long and 85% for short GRBs) could not be explained by a Maxwellian population-based slow-cooling synchrotron function. Our results show that using the time-resolved spectra this violation is actually more severe, with over 91% of spectra obtained from long bursts violating the Maxwellian synchrotron function drawn from a single temperature, which is already a limiting case.

Besides non-thermal models, one may construct a thermal emission dominated model to explain the observed sharpness angles. However, fitting multiple blackbodies is statistically meaningless, as one may construct any function from a polynomial fit. Simple photospheric models face difficulties in explaining the observed data. For example, early theoretical studies of a pure thermal origin of GRB prompt emission, such as from freely expanding photospheric outflows with no baryonic matter or magnetic field (Goodman 1986; Paczynski 1986), struggled to explain the shape of the prompt phase and the two modes of observed $E_p$ evolutionary trends (i.e., hard-to-soft evolution and intensity tracking, see, e.g., Ford et al. 1995). Recent studies (e.g., Pe'er et al 2006; Giannios 2008; Pe'er & Ryde 2011; Vurm et al. 2011; Ryde et al. 2011; Lazzati et al. 2013) suggested that the Band function can be reconstructed from a thermal model. However, Deng & Zhang (2014) claim that the hard-to-soft evolution of $E_p$ is difficult to reproduce under natural photospheric conditions.

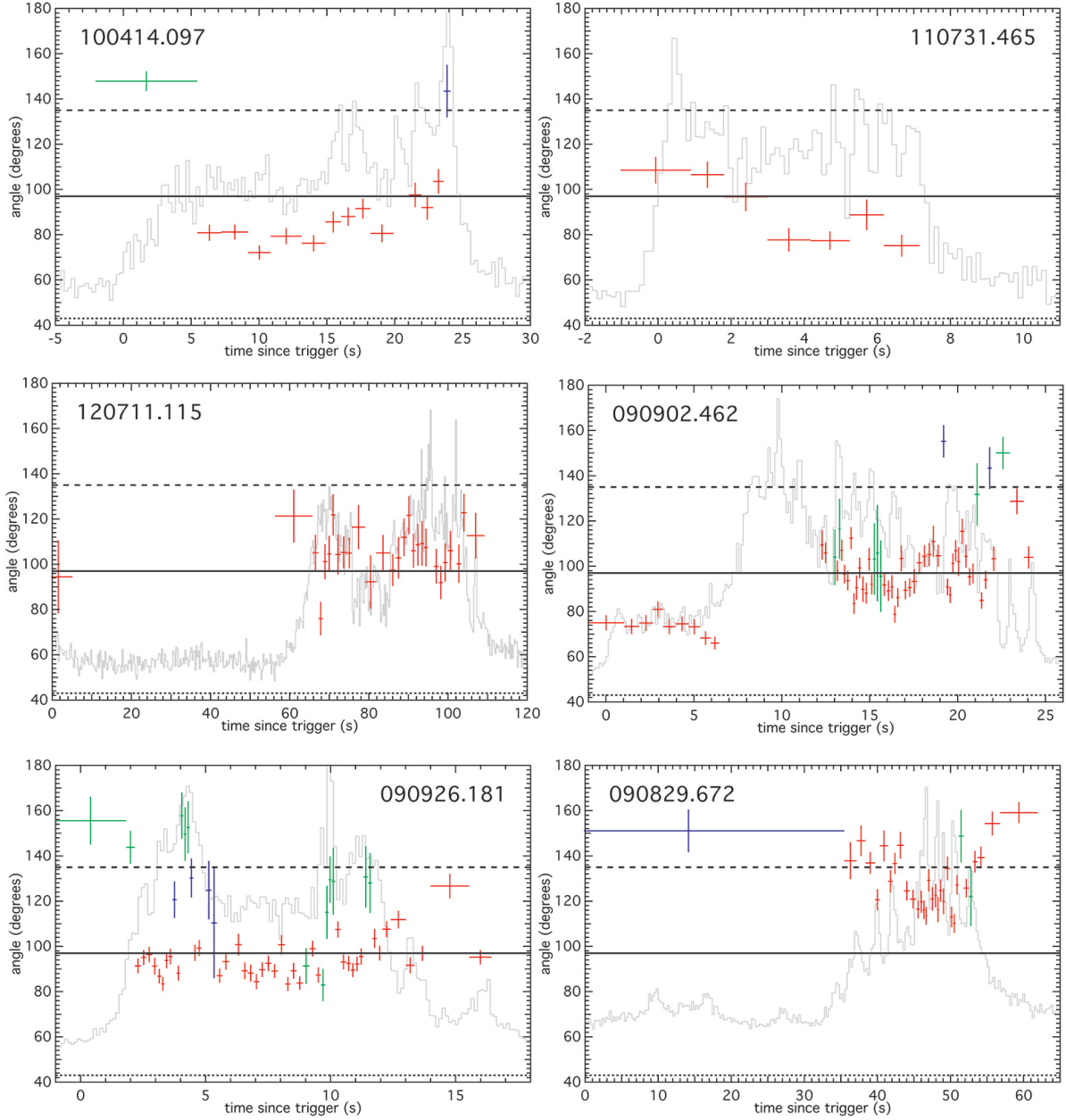

**Fig. 5.** Six examples of evolutionary trends of $\theta$. Red, blue, or green color indicates that the best-fit model is exponential cutoff power law (COMP), Band function (BAND), or smoothly broken power law (SBPL), respectively. The light curves are overlaid in arbitrary units. The limits of the normalized blackbody (dotted line), single-electron synchrotron (solid line), and synchrotron emission from a Maxwellian electron distribution (dashed line) are overlaid.

A frequently discussed alternative to the baryonic composition of the jets in GRBs is a magnetically, or Poynting flux, dominated jet (Thompson 1994; Drenkhahn & Spruit 2002; Lyutikov & Blandford 2003). In this scenario, the magnetic field dominates the energy density in the emitting region. Thus, the dominant emission mechanism will be synchrotron emission from relativistic electrons, since no cooling mechanism is known which is faster (see, e.g., Beniamini & Piran 2014). Our observational results therefore posed a challenge to Poynting flux dominated models, although Compton up-scattering from seed photons in the environment of an emerging Baryon-free jet offer a potential means of combining strongly magnetic outflows with a thermalized component or sharp spectrum (see, Gill & Thompson 2014, for a recent example). Moreover, Beloborodov (2013) argues that other optically thin emission models share the same problems of the synchrotron emission models, e.g., pitch-angle synchrotron radiation (Lloyd &

Petrosian 2000) when the scatter angle in the comoving frame is not isotropic, and jitter radiation in turbulent magnetic fields (Medvedev 2000).

The minimum variability timescale (MVT, e.g., Bhat 2013; Golkhou & Butler 2014) of the light curves provides dynamical timescale of the emission process. In Fig. 6, we plot $\theta$ against temporal bin widths per MVT (for the computational method of the MVT, see Bhat 2013). It is observed that, in 1,064 spectra (49 spectra were excluded because they belong to bursts with no MVT due to bad or not enough GBM data), only 4.4% of the spectra have bin width less than the MVT for the respective burst. This means that the problem for the synchrotron theory may be even more severe, since our spectra are smoothened already by integrating over multiple emitting shells.

To investigate the effect of integrating multiple spectra, we compute the average time-resolved sharpness angle $\langle\theta\rangle$ by weighing each spectrum equally. In Fig. 7, we compare $\langle\theta\rangle$ to the sharpness angle computed using the time-integrated catalog (Gruber et al. 2014), $\theta^{int}$, for every burst in our sample. Green color indicates the 7 bursts (10%) whose average sharpness angles are consistent with the Maxwellian synchrotron limit (note that individual $\theta$ values can still be inconsistent), orange color indicates the 55 bursts (79%) that are inconsistent with the Maxwellian synchrotron limit but consistent on average with the single-electron synchrotron limit, and red color indicates the 8 bursts (11%) that are inconsistent with the single-electron synchrotron limit. We note that the error bars of $\langle\theta\rangle$ represent the standard deviations, which indicate the spread of the angle distributions within each burst. The error bars of $\theta^{int}$ are computed using the same procedure as described in Section 3, and are relatively small because the parameters are better constrained by higher photon counts.

Fig. 7 shows that the time-integrated angles are systematically larger than the average time-resolved angles for individual bursts. We emphasize that different light curve binning methods are used in the time-resolved and time-integrated spectral catalog. In our time-resolved analysis, as mention in Yu et al. (2015b), the light curves are binned with $S/N = 30$, and then those spectra without a peak or break are excluded. In the time-integrated catalog (see, e.g., Gruber et al. 2014), all time intervals with $S/N \geq 3.5$ are included. The fact that fewer bursts in the time-integrated spectral analysis are inconsistent with the Maxwellian limit underlines the importance of time-resolved analysis.

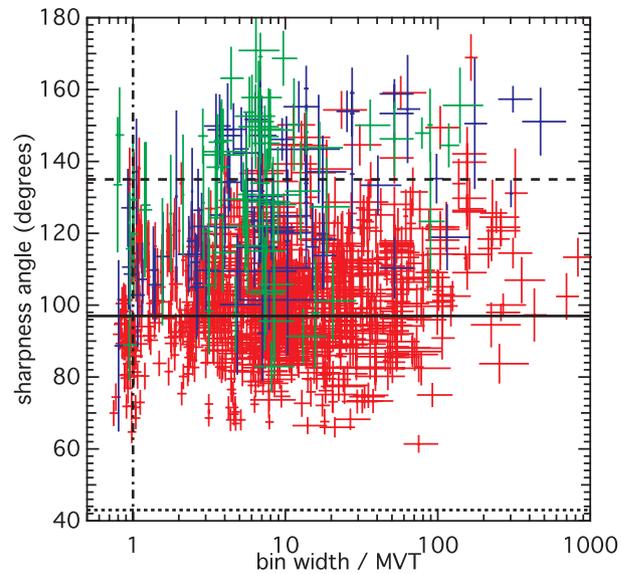

**Fig. 6.** Sharpness angles plotted against the temporal bin widths per MVT. Red data points show spectra best fit by the exponential cutoff power law (COMP), blue by the Band function (BAND), and green by the smoothly broken power law (SBPL). The vertical dash-dotted line shows where the bin width equals the MVT, only 4.4% of data points are located to the left of the line. The horizontal lines show the limits of the normalized blackbody (dotted), single-electron synchrotron (solid), and synchrotron emission from a Maxwellian electron distribution (dashed).

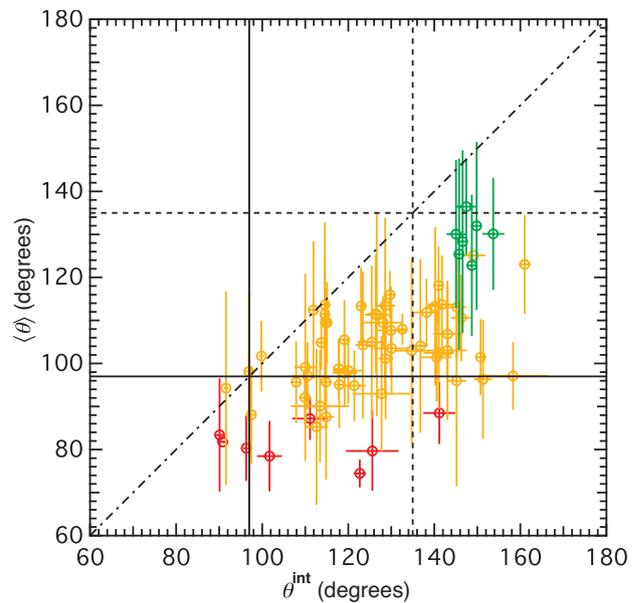

**Fig. 7.** Comparison between the average sharpness angles, $\langle\theta\rangle$, to the sharpness angles computed using the time-integrated catalog, $\theta^{int}$. The dash-dotted line shows $x = y$. The solid and dashed lines show the single-electron synchrotron and Maxwellian synchrotron limit, respectively. We note that the error bars of $\langle\theta\rangle$ represent the spread in $\theta$. See main text for the color-coding and details about the plots.

## 5. CONCLUSIONS

We reported the results given by Yu et al. (2015b) of the novel quantity, the sharpness angles $\theta$, obtained from the observed time-resolved spectra of *Fermi* GRBs. The values are compared to the sharpest cases of the synchrotron radiation theory, namely the single-electron synchrotron and the Maxwellian distributed synchrotron emission function. More than 91% of the observed spectra are found to be sharper than the Maxwellian synchrotron function, indicating that a simple synchrotron radiation mechanism cannot be responsible for the peaks or breaks of GRB prompt emission spectra. No general evolutionary trend is observed for $\theta$ within bursts. Moreover, the Maxwellian synchrotron function can only contribute up to $58^{+23}_{-18}\%$ of the peak flux. We conclude that the underlying prompt emission mechanism in GRBs must produce spectra sharper than a Maxwellian synchrotron function but broader than a blackbody.

It is still possible for synchrotron emission to dominate the spectrum away from the peak or break observed in the GBM energy range (e.g., at the LAT energy range). Also, a sub-dominant synchrotron component can allow for a continuous connection to the afterglow phase, where synchrotron emission is typically dominant (see, e.g., van Eerten 2015, for a recent review). The transition between prompt and afterglow is then marked by the disappearance of the non-synchrotron (likely thermal) component. There are other theoretical possibilities to explain GRB prompt emission, such as the collisional model of electron-positron pairs (e.g., Beloborodov 2010). For recent reviews on GRB prompt emission mechanisms, see, e.g., Zhang (2014) and Pe'er (2015).

A possibly similar inference can be made on the related phenomena of prompt optical emission showing a similar temporal profile as the gamma-ray emission (Elliott et al. 2014; Greiner et al. 2014) or very early X-ray flares (e.g., Pe'er et al. 2006; see also Hu et al. 2014 for a recent large *Swift* sample study): if the prompt emission is not dominated by synchrotron emission, this is likely the case for this longer wavelength emission as well (see, e.g., Starling et al. 2012; Peng et al. 2014).

Yu et al. (2015b) demonstrated a new method to quantify the shape of the observed GRB spectra, providing a tool for distinguishing between various standard emission functions. Ultimately, the question as to the viability of any particular emission model can only be fully resolved if complete spectral predictions for that model are tested directly against photon counts (see, e.g., Burgess et al. 2014).

## ACKNOWLEDGMENTS

HFY and JG acknowledge support by the DFG cluster of excellence 'Origin and Structure of the Universe' (www.universe-cluster.de). HJvE acknowledges support by the Alexander von Humboldt foundation. RS is partially supported by ISF, ISA and iCore grants. The GBM project is supported by the German Bundesministeriums für Wirtschaft und Technologie (BMWi) via the Deutsches Zentrum für Luft und Raumfahrt (DLR) under the contract numbers 50 QV 0301 and 50 OG 0502.